\documentclass[conference]{IEEEtran}
\usepackage{cite}
\usepackage{amsmath}
\usepackage{algorithmic}
\usepackage{array}
\usepackage{fixltx2e}
\usepackage{stfloats}
\usepackage{url}
\usepackage{graphicx}
\usepackage{algorithmic}
\usepackage{algorithm}
\usepackage{tabularx}
\usepackage[colorlinks=true, linkcolor=blue, citecolor=blue, urlcolor=blue]{hyperref}
\begin{document}

\title{LLM-Enabled EV Charging Stations Recommendation}

\author{\IEEEauthorblockN{Zeinab Teimoori}
\IEEEauthorblockA{Department of Engineering, Thompson Rivers University, Canada\\
Email: zteimoori@tru.ca}
}

\maketitle

\begin{abstract}
Charging infrastructure is not expanding quickly enough to accommodate the increasing usage of Electric Vehicles (EVs). For this reason, EV owners experience extended waiting periods, range anxiety, and overall dissatisfaction. Challenges, such as fragmented data and the complexity of integrating factors like location, energy pricing, and user preferences, make the current recommendation systems ineffective. To overcome these limitations, we propose \textit{RecomBot}, which is a Large Language Model (LLM)-powered prompt-based recommender system that dynamically suggests optimal Charging Stations (CSs) using real-time heterogeneous data. By leveraging natural language reasoning and fine-tuning EV-specific datasets, \textit{RecomBot} enhances personalization, improves charging efficiency, and adapts to various EV types, offering a scalable solution for intelligent EV recommendation systems. Through testing across various prompt engineering scenarios, the results obtained underline the capability and efficiency of the proposed model.
\end{abstract}


\IEEEpeerreviewmaketitle

\section{Introduction}
\begin{figure}
  \centering
  \includegraphics[width=\columnwidth, height = 170pt]{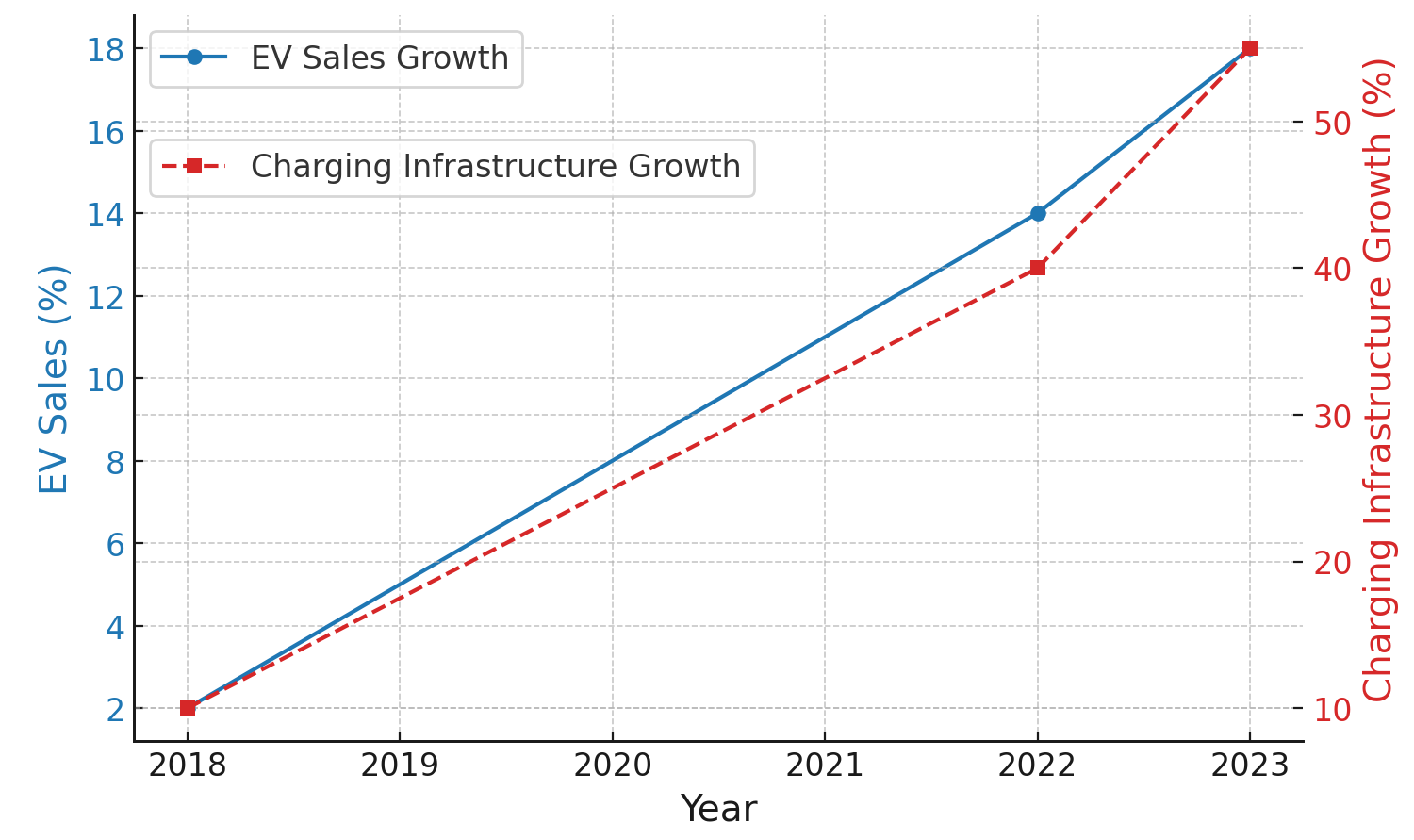}
  \caption{Insufficient growth of EV charging infrastructure vs. EV sales \cite{EV_Outlook}.}
  \label{EV_CS_growth}
\end{figure}

The fast-growing use of Electric Vehicles (EVs) is essential for advancing the transition to a more sustainable transportation system. Governments and industries are investing heavily in EV technology to promote policies that encourage electrification and to expand charging infrastructure to satisfy the increasing demand. However, as the number of EV users grows, so does the demand for dependable and easily accessible Charging Stations (CSs) \cite{jin2024democratizing, TeimooriS:2022}. Figure \ref{EV_CS_growth} visualizes the insufficient growth of EV charging infrastructure compared to EV sales growth, which, despite significant expansion, still struggles to keep up with the accelerating request for EVs. The efficiency of the charging structure is notable for supporting user satisfaction and the ongoing growth of EV adoption. \cite{yousuf2024depth}. Inadequate or inefficiently managed CSs can lead to long wait times, range anxiety, and reduced trust in EV technology. If users struggle to find suitable CSs, especially in high-traffic areas or adverse weather conditions, the benefits of EV adoption could be surpassed by infrastructure limitations \cite{yuvaraj2024comprehensive, TeimooriCon:2024}.

To enhance user experience and optimize CS utilization, user-centric recommendation systems are essential. However, there are several challenges, such as data scarcity, as real-time CS information is often fragmented across multiple providers and may not be consistently updated \cite{algafri2024smart, TeimooriRec:2024}. Furthermore, EV charging involves heterogeneous data sources, including time-series data, geographical data, weather conditions, energy grid load, and user preferences. Understanding and integrating these diverse data types into a meaningful recommendation model requires refined processing and reasoning capabilities \cite{teimoori2022secure}.

Past progress in Large Language Models (LLMs) has shown considerable improvement in addressing complicated problems across various domains \cite{mongaillard2024large}, including the EV industry \cite{patil2024towards, qu2024chatev, feng2024large}. In this paper, we propose a novel prompt-based recommender system \textit{(RecomBot)} that leverages LLMs to generate a dynamic list of the best CS matches for users. By integrating multi-modal data with natural language reasoning, with their ability to process large-scale data, analyze user preferences, and generate optimal suggestions, our framework improves the efficiency, accessibility, and dependability of the EV charging foundation.

The main contributions of our proposed model are outlined below:
\begin{itemize}
    \item Develop \textit{RecomBot}, an LLM-powered chatbot that recommends CSs based on real-time data adaptability to dynamic charging station conditions.
    \item Incorporate multi-modal data processing (energy pricing, Google reviews, user preferences) to enable proactive context-aware optimized suggestions.
    \item Enhance personalization by fine-tuning the LLM on EV-specific user-centric data to improve charging efficiency, reduce wait times, and enhance the user experience.
\end{itemize}

The following sections are structured as follows: A comprehensive explanation of the proposed system is presented in Section \ref{method}. Then the analysis of the findings is presented in Section \ref{eva}. Finally,  Section \ref{con} concludes the key insights.

\bigskip
\section{Methodology} \label{method}

\begin{figure*}
  \centering
  \includegraphics[width=380pt, height=270pt]{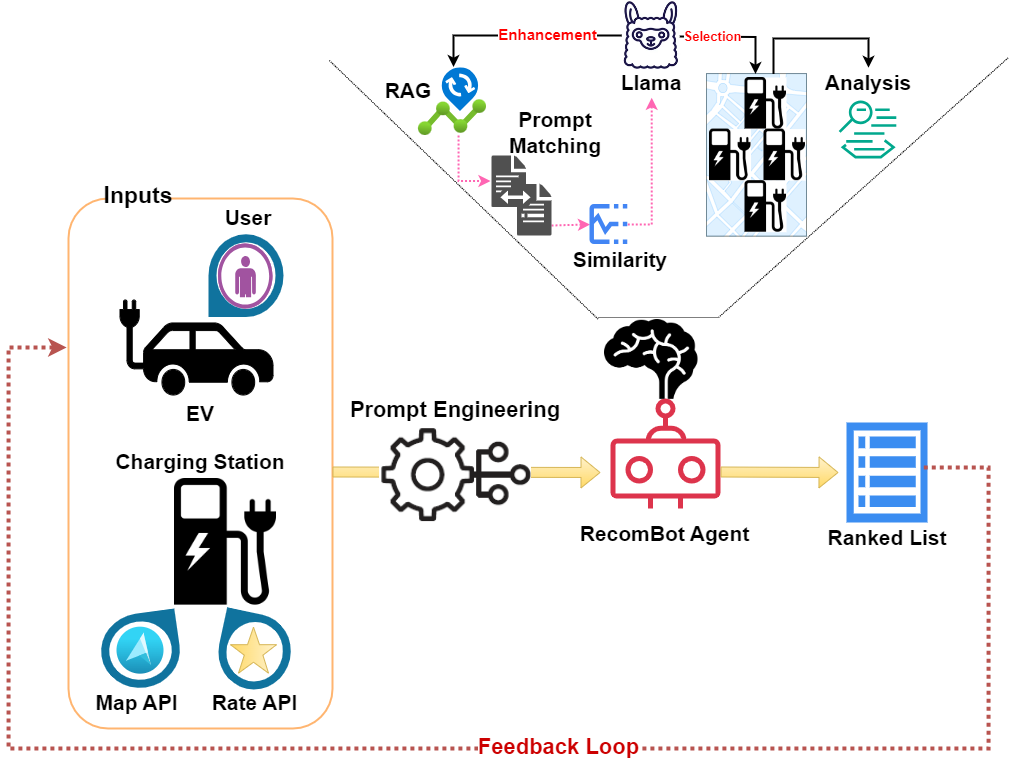}
  \caption{\textit{RecomBot} agent framework integrates prompt engineering and optimization techniques to refine selections based on user preferences.}
  \label{framework}
\end{figure*}

When developing recommendation methods powered by Artificial Intelligence (AI), it is important to acknowledge that conventional methods are usually built to function within fixed constraints. These constraints are inherently linked to the system's architecture and goals, which can restrict their flexibility and ability to generalize across different contexts \cite{patil2024towards, zhang2025advancing}. In this article, we propose a framework (\textit{RecomBot}) that leverages LLMs (LLaMA 3 8B model) to process user queries and generate optimized EV charging recommendations. The system integrates multiple data sources, applies prompt engineering for LLM-based reasoning, and outputs ranked CS recommendations based on user preferences. The general architecture is illustrated in Figure \ref{framework}. As shown in Figure \ref{framework}, this model follows several steps to generate the desired results. Here in the rest of this section, we explain each step in detail.

\bigskip
\noindent\textbf{Step one: User Query Processing and Preference Normalization}

Initially, an EV user submits a natural language query \( Q \) specifying charging constraints such as location, cost, power, or station availability ("Find me the nearest fast charger under \$0.50/kWh"). \textit{RecomBot} extracts key preference parameters using intent recognition. Let:
\[ Q = \{q_1, q_2, ..., q_n\}, \]
where each \( q_i \) represents a distinct preference. The preference weight vector, which is based on the user's request and priority, is defined as:
\[ \mathbf{w} = [w_\text{distance}, w_\text{price}, w_\text{power}, w_\text{rating}, ...], \]
\bigskip

and normalized before matching to predefined categories, such that:

\begin{equation}
    \sum_{i=1}^{n} w_i = 1.
\end{equation}

\bigskip
\noindent\textbf{Step two: Real-Time Data Integration and Retrieval Augmented Generation (RAG)}

To fetch relevant CS data, the framework calls external APIs (Application Programming Interface) such as the Open Charge Map API \cite{openchargemap}, and the Google Cloud API for station information and energy price databases. The retrieved station set is denoted as:
\[ S = \{s_1, s_2, ..., s_m\}, \]
where each station \( s_i \) has attributes \( (d_i, p_i, r_i, c_i) \) representing distance, price, rating, and charging speed, respectively.

The RAG module inside the model enhances the \textit{RecomBot}'s reasoning by retrieving relevant CS details, and then the similarity between a station \( s_i \) and the user’s preference vector is computed using cosine similarity \cite{mongaillard2024large}:

\begin{equation}
    \text{sim}(Q, s_i) = \frac{\mathbf{w} \cdot \mathbf{s}_i}{||\mathbf{w}|| ||\mathbf{s}_i||}. 
\end{equation}

\bigskip
\noindent\textbf{Step three: Optimization Problem Formulation}

\textit{RecomBot} converts the user request into a constrained optimization problem to maximize the relevance score for each CS, as shown below:
\begin{equation}
    \max_{s_i \in S} \sum_{j=1}^{n} w_j s_{i,j},
\end{equation}

\noindent subject to: 
\[ d_i \leq d_\text{max}, \quad p_i \leq p_\text{max}, \quad c_i \geq c_\text{min}, \] 

\noindent where \( d_\text{max} \), \( p_\text{max} \), and \( c_\text{min} \) are user-defined constraints.

\bigskip
\noindent\textbf{Step four: Ranked Recommendation Generation}

Once relevant stations are retrieved, \textit{RecomBot} formats the data into structured prompts for LLM processing. The final recommendation list is generated as:
\begin{equation}
    R = \text{sort}(S, \text{sim}(Q, s_i)),
\end{equation}

\noindent and presented to the user with an explanation of the ranking rationale ("I recommended Station X because it has the lowest price and shortest distance.").

\bigskip
\noindent\textbf{Step five: Feedback Loop and Adaptive Learning}

User feedback \( f \) is gathered after the interaction, and used to adjust preference weights via reinforcement learning \cite{feng2024large}:
\begin{equation}
    w_j^{(t+1)} = w_j^{(t)} + \eta f_j,
\end{equation}
where \( \eta \) is the learning rate. This adaptive mechanism improves \textit{RecomBot}'s performance over time.

\bigskip
\section{Evaluation and Results} \label{eva}
\begin{figure}
  \centering
  \includegraphics[width=\columnwidth, height=200pt]{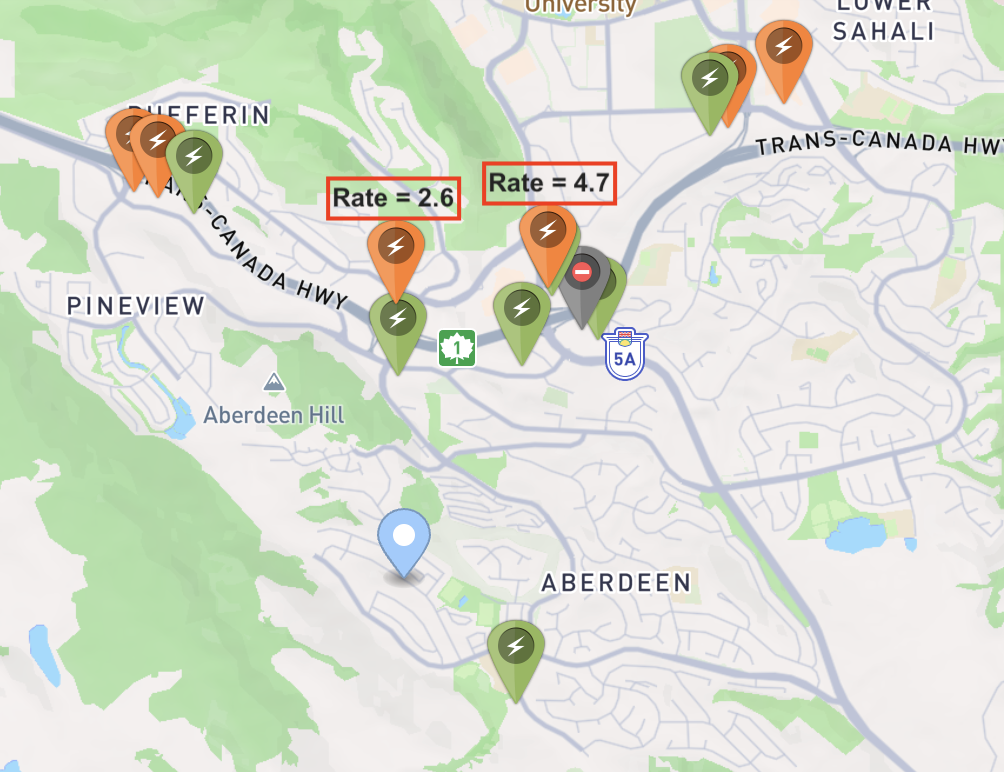}
  \caption{Station retrieved from the API surrounding the EV user location (in blue).}
  \label{map}
\end{figure}

To evaluate the performance of the proposed system, we tested \textit{RecomBot} in Kamloops, Canada. The evaluation considers a user located at latitude 50.640054 and longitude -120.378926. Figure \ref{map} illustrates the identified CSs in the vicinity of the user location. The system correctly retrieves stations within a reasonable radius and categorizes them based on their attributes. Orange markers indicate fast stations, while green markers highlight lower-powered ones. 

As explained before, \textit{RecomBot} ranks the retrieved charging stations using user-defined preferences, such as (1) Distance: Prioritizing stations that are closest to the user's current location, (2) Price: Selecting cost-effective charging options, (3) Power Output: Recommending high-power stations for faster charging, and (4) Rating: Preferring well-rated stations to ensure user satisfaction.

\begin{figure*}
  \centering
  \includegraphics[height=400pt, width=\textwidth]{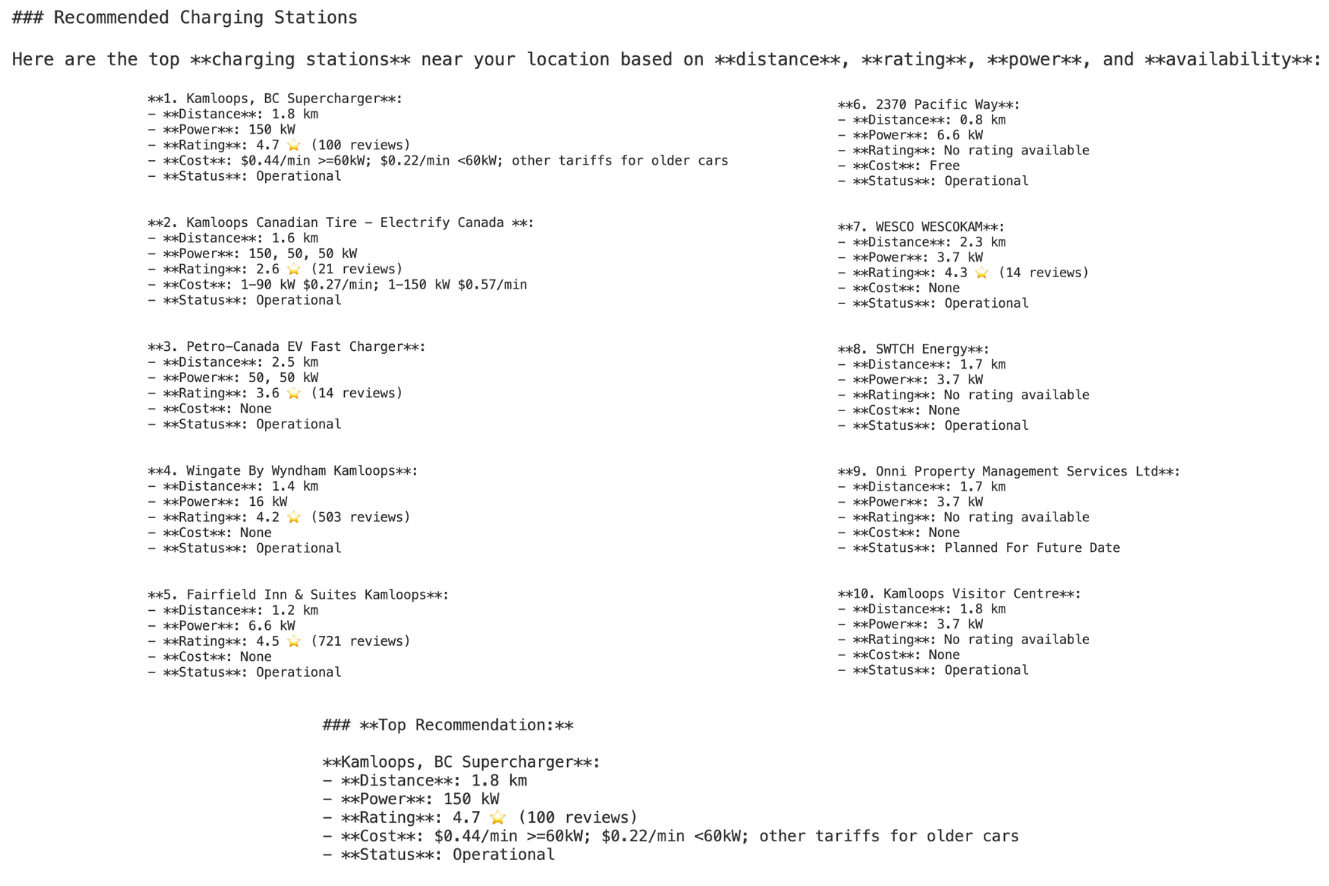}
  \caption{\textit{RecomBot} performance based on the user preference "\textit{fast charging high rated}".}
  \label{RecomBot}
\end{figure*}

The recommendation list, as shown in Figure \ref{RecomBot}, highlights the top-ranked CSs based on the user prompt \textit{"fast charging high rated"}. The highest-rated CS (Kamloops, BC Supercharger) has a power output of 150 kW and a rating of 4.7. In contrast, the second option, Kamloops Canadian Tire-Electrify Canada, has a rating of 2.6 but supports fast charging with a power output of 150 kW.

\textit{RecomBot} dynamically adjusts rankings based on user preferences. For example, if the user requests \textit{"fast charging near me"}, the system prioritizes power and distance. If the user asks for \textit{"cheap high-rated stations"}, the system optimizes based on price and rating. If multiple preferences are combined, such as \textit{"fast charging cheap high-rated"}, the system balances power, price, and rating in its rankings to aim for the optimized suggestion. Tables \ref{rank_a} and \ref{rank_b} show how \textit{RecomBot} ranks CSs based on different user preferences.

\bigskip
\begin{table}[h]
    \centering
    \caption{User Preference "fast charging near me" }
    \begin{tabular}{cllc}
         \vspace{0.3em}
         \textbf{Rank } & \textbf{Station Name} & \textbf{Power Level} & \textbf{Distance} \\
         \hline
         1 & Kamloops Canadian Tire & 150 kW & 1.6 km\\ \vspace{0.2em}
         2 & Kamloops Supercharger & 150 kW & 1.8 km\\ \vspace{0.2em}
         3 & Fairfield Inn & 6.6 kW & 1.2 km\\
    \end{tabular}
    \label{rank_a}
\end{table}

\bigskip
\begin{table}[h]
    \centering
    \caption{User Preference "cheap high rating" }
   \begin{tabular}{cllll}
         \vspace{0.3em}
         \textbf{Rank } & \textbf{Station Name} & \textbf{Power Level} & \textbf{Cost} & \textbf{Rating} \\
         \hline \vspace{0.2em}
         1 & 2370 Pacific Way & 6.6 kW & Free & 4.5 $\star$  \\ \vspace{0.2em}
         2 & Fairfield Inn & 6.6 kW & Free & 4.5 $\star$\\ \vspace{0.2em}
         3 & Kamloops Supercharger & 150 kW & \$0.44/min & 4.7 $\star$\\
    \end{tabular}
    \label{rank_b}
\end{table}

\pagebreak
These evaluations demonstrate that \textit{RecomBot} efficiently processes user queries, retrieves relevant CSs, and provides personalized recommendations.

\bigskip
\section{Conclusion and Future Work} \label{con}
In this work we introduced an LLM-powered recommendation system, \textit{RecomBot}, to optimize EV CS selection through RAG, optimization modeling, and reinforcement learning. By incorporating data fusion, it integrated structured and unstructured data to handle ambiguous user queries for a personalized and efficient charging experience. \textit{RecomBot} enhanced EV navigation by leveraging multi-modal data processing and adaptive learning, which enables scalability and flexibility for various EV types, including bikes, taxis, buses, and fleet stations. Future extensions include integrating mobile CSs when fixed stations are crowded and incorporating weather and traffic conditions to improve recommendations.

\bigskip
\section*{Acknowledgment}
Partial support for this work was provided by the Thompson Rivers University Research Office.

\bigskip
\bibliographystyle{IEEEtran}
\bibliography{./Main}

\begin{thebibliography}{10}
\providecommand{\url}[1]{#1}
\csname url@samestyle\endcsname
\providecommand{\newblock}{\relax}
\providecommand{\bibinfo}[2]{#2}
\providecommand{\BIBentrySTDinterwordspacing}{\spaceskip=0pt\relax}
\providecommand{\BIBentryALTinterwordstretchfactor}{4}
\providecommand{\BIBentryALTinterwordspacing}{\spaceskip=\fontdimen2\font plus
\BIBentryALTinterwordstretchfactor\fontdimen3\font minus \fontdimen4\font\relax}
\providecommand{\BIBforeignlanguage}[2]{{%
\expandafter\ifx\csname l@#1\endcsname\relax
\typeout{** WARNING: IEEEtran.bst: No hyphenation pattern has been}%
\typeout{** loaded for the language `#1'. Using the pattern for}%
\typeout{** the default language instead.}%
\else
\language=\csname l@#1\endcsname
\fi
#2}}
\providecommand{\BIBdecl}{\relax}
\BIBdecl

\bibitem{jin2024democratizing}
M.~Jin, B.~Sel, F.~Hardeep, and W.~Yin, ``Democratizing energy management with llm-assisted optimization autoformalism,'' in \emph{2024 IEEE International Conference on Communications, Control, and Computing Technologies for Smart Grids (SmartGridComm)}.\hskip 1em plus 0.5em minus 0.4em\relax IEEE, 2024, pp. 258--263.

\bibitem{TeimooriS:2022}
\BIBentryALTinterwordspacing
Z.~Teimoori and A.~Yassine, ``A review on intelligent energy management systems for future electric vehicle transportation,'' \emph{Sustainability}, vol.~14, no.~21, 2022. [Online]. Available: \url{https://www.mdpi.com/2071-1050/14/21/14100}
\BIBentrySTDinterwordspacing

\bibitem{yousuf2024depth}
A.~Yousuf, Z.~Wang, R.~Paranjape, and Y.~Tang, ``An in-depth exploration of electric vehicle charging station infrastructure: a comprehensive review of challenges, mitigation approaches, and optimization strategies,'' \emph{IEEE Access}, 2024.

\bibitem{yuvaraj2024comprehensive}
T.~Yuvaraj, K.~Devabalaji, J.~A. Kumar, S.~B. Thanikanti, and N.~I. Nwulu, ``A comprehensive review and analysis of the allocation of electric vehicle charging stations in distribution networks,'' \emph{IEEE Access}, vol.~12, pp. 5404--5461, 2024.

\bibitem{TeimooriCon:2024}
Z.~Teimoori and A.~Yassine, ``User-centric charging service recommendation for electric vehicles,'' in \emph{2024 IEEE 22nd Mediterranean Electrotechnical Conference (MELECON)}, 2024, pp. 739--743.

\bibitem{EV_Outlook}
{iea.org}, ``Global ev outlook 2024,'' Available online: \url{https://www.iea.org/reports/global-ev-outlook-2024}, (accessed on 02 2025).

\bibitem{algafri2024smart}
M.~Algafri, A.~Alghazi, Y.~Almoghathawi, H.~Saleh, and K.~Al-Shareef, ``Smart city charging station allocation for electric vehicles using analytic hierarchy process and multiobjective goal-programming,'' \emph{Applied Energy}, vol. 372, p. 123775, 2024.

\bibitem{TeimooriRec:2024}
Z.~Teimoori, A.~Yassine, and M.~S. Hossain, ``Smart vehicles recommendation system for artificial intelligence-enabled communication,'' \emph{IEEE Transactions on Consumer Electronics}, vol.~70, no.~1, pp. 3914--3925, 2024.

\bibitem{teimoori2022secure}
------, ``A secure cloudlet-based charging station recommendation for electric vehicles empowered by federated learning,'' \emph{IEEE Transactions on Industrial Informatics}, vol.~18, no.~9, pp. 6464--6473, 2022.

\bibitem{mongaillard2024large}
T.~Mongaillard, S.~Lasaulce, O.~Hicheur, C.~Zhang, L.~Bariah, V.~S. Varma, H.~Zou, Q.~Zhao, and M.~Debbah, ``Large language models for power scheduling: A user-centric approach,'' in \emph{2024 22nd International Symposium on Modeling and Optimization in Mobile, Ad Hoc, and Wireless Networks (WiOpt)}.\hskip 1em plus 0.5em minus 0.4em\relax IEEE, 2024, pp. 321--328.

\bibitem{patil2024towards}
M.~S. Patil, G.~Ung, and M.~Nyberg, ``Towards specification-driven llm-based generation of embedded automotive software,'' in \emph{International Conference on Bridging the Gap between AI and Reality}.\hskip 1em plus 0.5em minus 0.4em\relax Springer, 2024, pp. 125--144.

\bibitem{qu2024chatev}
H.~Qu, H.~Li, L.~You, R.~Zhu, J.~Yan, P.~Santi, C.~Ratti, and C.~Yuen, ``Chatev: Predicting electric vehicle charging demand as natural language processing,'' \emph{Transportation Research Part D: Transport and Environment}, vol. 136, p. 104470, 2024.

\bibitem{feng2024large}
J.~Feng, C.~Cui, C.~Zhang, and Z.~Fan, ``Large language model based agent framework for electric vehicle charging behavior simulation,'' \emph{arXiv preprint arXiv:2408.05233}, 2024.

\bibitem{zhang2025advancing}
H.~Zhang, R.~Zhang, W.~Zhang, D.~Niyato, and Y.~Wen, ``Advancing generative artificial intelligence and large language models for demand side management with electric vehicles,'' \emph{arXiv preprint arXiv:2501.15544}, 2025.

\bibitem{openchargemap}
{Open Charge Maps}, ``Locating charging stations,'' Available online: \url{https://openchargemap.org/site}, (accessed on 02 2025).

\end{thebibliography}
\end{document}